\documentclass[a4paper]{article}
\usepackage[bottom]{footmisc}
\usepackage{INTERSPEECH2021}
\usepackage{multirow}
\usepackage[colorlinks,linkcolor=black]{hyperref}
\title{AISHELL-4: An Open Source Dataset for Speech Enhancement, Separation, Recognition and Speaker Diarization in Conference Scenario}
\name{Yihui Fu$^{1,*}$, Luyao Cheng$^{1,*}$\thanks{* Contribute equally}, Shubo Lv$^1$, Yukai Jv$^1$, Yuxiang Kong$^1$, Zhuo Chen$^2$ \\ Yanxin Hu$^1$, Lei Xie$^{1,\#}$\thanks{\# Corresponding author}, Jian Wu$^3$, Hui Bu$^4$, Xin Xu$^4$, Jun Du$^5$, Jingdong Chen$^1$}
\address{
	$^1$Northwestern Polytechnical University, Xi'an, China\\
	$^2$Microsoft Corporation, USA\\
	$^3$Microsoft Corporation, China\\
	$^4$Beijing Shell Shell Technology Co., Ltd., Beijing, China\\
	$^5$University of Science and Technology of China, Hefei, China
}
\email{\{yhfu, lycheng, shblv, ykjv, yxkong, yxhu\}@npu-aslp.org, lxie@nwpu.edu.cn, \\ \{zhuc, wujian\}@microsoft.com, \{buhui, xuxin\}@aishelldata.com, jundu@ustc.edu.cn, jingdongchen@ieee.org}
\begin{document}
	
	\maketitle
	\begin{abstract}
		In this paper, we present AISHELL-4, a sizable real-recorded Mandarin speech dataset collected by 8-channel circular microphone array for speech processing in conference scenario. The dataset consists of 211 recorded meeting sessions, each containing 4 to 8 speakers, with a total length of 120 hours. 
		This dataset aims to bridge the advanced research on multi-speaker processing and the practical application scenario in three aspects. With real recorded meetings, AISHELL-4 provides realistic acoustics and rich natural speech characteristics in conversation such as short pause, speech overlap, quick speaker turn, noise, etc. 
		Meanwhile, accurate transcription and speaker voice activity are provided for each meeting in AISHELL-4. This allows the researchers to explore different aspects in meeting processing, ranging from individual tasks such as speech front-end processing, speech recognition and speaker diarization, to multi-modality modeling and joint optimization of relevant tasks.
		Given most open source dataset for multi-speaker tasks are in English, AISHELL-4 is the only Mandarin dataset for conversation speech, providing additional value for data diversity in speech community.
		We also release a PyTorch-based training and evaluation framework as baseline system to promote reproducible research in this field.
	\end{abstract}
	\noindent\textbf{Index Terms}: AISHELL-4, speech front-end processing, speech recognition, speaker diarization, conference scenario, Mandarin
	
	\section{Introduction}
	
	Meeting transcription defines a process of estimating ``who speaks what at when'' on meetings recordings, which usually composes of utterances from multiple speakers, with a certain amount of speech overlap.
	It is considered as one of the most challenging problems in speech processing, as it is a combination of various speech tasks, including speech front-end processing, speech activity detection (SAD), automatic speech recognition (ASR), speaker identification and diarization, etc. While several works show promising results~\cite{kanda2021investigation, wang2021exploring}, the problem is still considered unsolved, especially for real world applications~\cite{yoshioka2019advances}. 
	
	The scarcity of relevant data is one of the main factors that hinders the development of advanced meeting transcription system. As meeting transcription involves multiple aspects of the input audio such as speech content, speaker identifications and their onset/offset time, precise annotation on all aspects is expensive and usually difficult to obtain. Meanwhile, meeting often contains noticeable amount of speech overlap, quick speaker turn, non-grammatical presentation and noise, further increasing the transcription difficulty and cost. For example, to transcribe the overlapped regions in a meeting, the transcriber is often required to listen to the audio multiple times in order to understand each involved speaker. Therefore, most of the previous works on meeting transcription or relevant tasks such as overlapped speech recognition employ synthetic datasets for experimentation~\cite{li2021dual, chen2020don, chen2020continuous}.

	To help with the data insufficiency problem, several relevant datasets have been collected and released~\cite{chen2020continuous,watanabe2020chime,janin2003icsi,mostefa2007chil,renals2008interpretation}. 
	However, most of the existing datasets suffer from various limitations, ranging from corpus setup such as corpus size, speaker \& spatial variety, collection condition, etc., to corpus content such as recording quality, accented speech, speaking style, etc.
	Meanwhile, almost all public available meeting corpus are based on English, which largely limits the data variation for meeting transcription systems. As each language possesses unique properties, the solution for English based meeting is likely to be sub-optimum for other languages. 
	
	In this work, in order to boost the research on advanced meeting transcription, we design and release a sizeable collection of real meetings, namely AISHELL-4\footnote{http://www.aishelltech.com/aishell\_4} dataset. AISHELL-4 is a Mandarian based corpus, containing 120 hours of meeting recording, using an 8-channel microphone array.
	The corpus is designed to cover a variety of aspects in real world meetings, including diverse recording conditions, various numbers of meeting participants and overlapped ratios. High quality transcriptions on multiple aspects are provided for each meeting sample, allowing the researcher to explore different aspects of meeting processing. A baseline system is also released along with the corpus, to further facilitate the research in this complicated problem.

	

	
	
	\section{Previous Works}
	
	%
	%

	For ASR with conversation speech, switchboard~\cite{godfrey1992switchboard}, collected by Texas Instruments, is a dataset of telephone communication with about 2,500 conversations by 500 native speakers. However, the data scale is relatively limited. Fisher~\cite{cieri2004fisher} is another conversational telephone speech dataset which has yielded more than 16,000 English conversations and 2,000 hours of speech in duration, covering various speaker and speaking styles. However, fisher, together with switchboard, is recorded with a sampling rate of 8 kHz, which does not meet the modern requirement of high accuracy meeting transcription. Furthermore, these datasets are recorded in single-speaker manner and do not contain speech overlapping scenario.
	
	To cope with the speech overlapping, a variety of datasets have been developed. WSJ0-2mix~\cite{hershey2016deep}, and its variants WHAM!~\cite{wichern2019wham} and WHAMR!~\cite{maciejewski2020whamr} are mostly adopted for speech separation evaluation, which consists of artificially mixed speech samples using Wall Street Journal dataset. 
	LibriMix~\cite{cosentino2020librimix}, derived from LibriSpeech~\cite{panayotov2015librispeech}, is an alternative to datasets mentioned above for noisy speech separation task. Its sizeable data scale introduces a higher generalization to model training. 
	However, all datasets mentioned above are synthetic data and are created in short-segmented fully overlapped manner, with severe mismatches with real conversation that the overlap ratio is usually below 20\% ~\cite{ccetin2006analysis}. 
	
	LibriCSS and CHiME-6 are recently introduced datasets for conference or indoor conversation transcription, with consideration of overlapped speech. 
	LibriCSS mixes the LibriSpeech utterances from different speakers to form single channel continuous utterances, which are then played by multiple Hi-Fi loudspeakers and recorded by a 7-channel microphone array in one conference venue. However, LibriSpeech is a reading style corpus with relatively stable speed and pronunciation, which mismatches with natural conversation. 
	On the other hand, the CHiME-6 challenge includes a track aimed at recognizing unsegmented dinner-party conversations recorded on multiple microphone arrays. However, the recording quality of CHiME-6 is relatively low, as a result of the existence of audio clipping, a large amount of interjection and laugh, signal attenuation, etc. The number of the participant for each CHiME-6 meeting is fixed to four, making it difficult to evaluate the generalization of meeting transcription systems. 
	ICSI~\cite{janin2003icsi}, CHIL~\cite{mostefa2007chil} and AMI~\cite{renals2008interpretation} are meeting corpus collected from academic laboratories, instead of commercial microphones, which have relatively limited speaker variation and restrict the universality of data.
	The data scale of Santa Barbara Corpus of Spoken American English~\cite{stupakov2012design} is considerable. Unfortunately, it only contains single-channel data.

	For Mandarin speech corpus, the commonly used open source datasets for speech recognition include AISHELL-1~\cite{aishell_2017}, ASHELL-2~\cite{du2018aishell}, aidatatang\footnote{http://www.openslr.org/62/}, etc. However, most of the datasets are recorded in near-field scenario and have no speech overlap and obvious noise and reverberation. To the best of our knowledge, there is no public available meeting dataset in Mandarin. 
	\section{Datasets}
	
	
	
	AISHELL-4 contains 120 hours of speech data in total, divided into 107.50 hours training and 12.72 hours evaluation set. Training and evaluation set contain 191 and 20 sessions, respectively. Each session consists of a 30-minute discussion by a group of participants. The total number of participants in training and evaluation sets is 36 and 25, with balanced gender coverage.
	
\begin{figure}[htb]
		\centering
		\includegraphics[width=0.40 \textwidth]{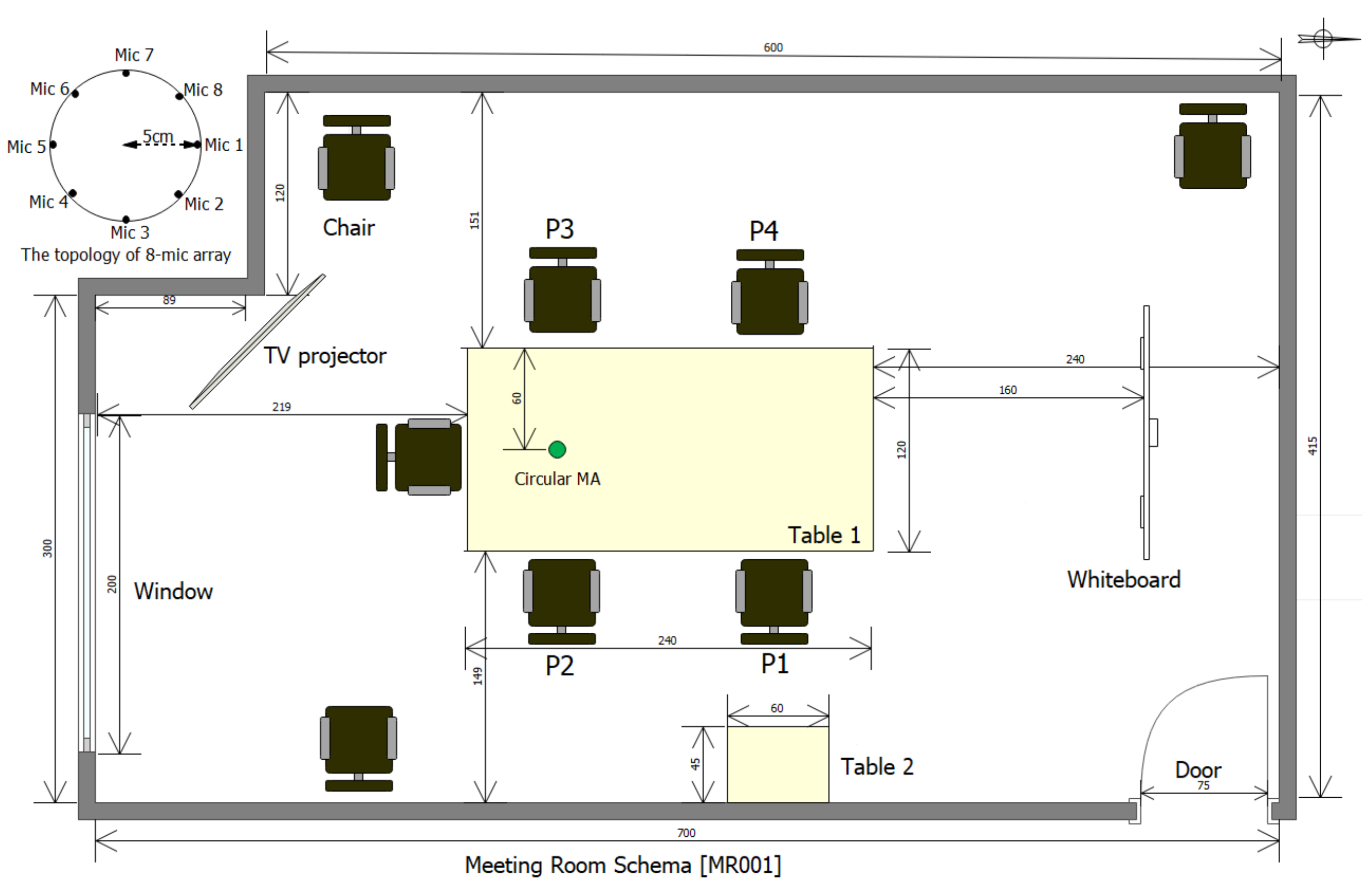}
		\caption{An example of recording venue of training set and the topology of microphone array.}
		\label{fig:topo}
	\end{figure}

	The dataset is collected in 10 conference venues.
	The conference venues are divided into three types: small, medium, and large room, whose size range from 7 $\times$ 3 $\times$ 3 to 15 $\times$ 7 $\times$ 3 m$^{3}$. The type of wall material of the conference venues covers cement, glass, etc. Other furnishings in conference venues include sofa, TV, blackboard, fan, air conditioner, plants, etc. During recording, the participants of the conference sit around the microphone array which is placed on the table in the middle of the room and conduct natural conversation. The microphone-speaker distance ranges from 0.6 to 6.0 m. All participants are native Chinese speakers speaking Mandarin without strong accents. During the conference, various kinds of indoor noise including but not limited to clicking, keyboard, door opening/closing, fan, bubble noise, etc., are made by participants naturally. For training set, the participants are required to remain in the same position during recording, while for evaluation set, the participants may move naturally within a small range. There is no room overlap and only one speaker overlap between training and evaluation set. An example of the recording venue from the training set, including the topology of microphone array, is shown in Fig. \ref{fig:topo}.
	
	\begin{table}[!htb]
		\caption{Details of AISHELL-4 dataset.}
		
		\label{table:data}
		\centering
		\begin{tabular}{c|c|c}
			\toprule
			\hline
			& Training & Evaluation \\ 
			\hline
			Duration (h) & 107.50 & 12.72 \\
			\#Session & 191 & 20 \\
			\#Room & 5 & 5 \\
			\#Participant & 36 & 25 \\
			\#Male & 16 & 11 \\
			\#Female & 20 & 14 \\
			Overlap Ratio (Avg.) & 19.04\% & 9.31\% \\
			
			\hline
			\bottomrule
		\end{tabular}
	\end{table}
	
	The number of participants within one conference session ranges from 4 to 8. To ensure the coverage of different overlap ratios, we select various meeting topics during recording, including medical treatment, education, business, organization management, industrial production and other daily routine meetings. The average speech overlap ratios of training and evaluation sets are 19.04~\% and 9.31~\%, respectively. More details of AISHELL-4 is shown in Table \ref{table:data}.
	A detailed session-level overlap ratio distribution of training and evaluation sets is shown in Table \ref{table:distribution}.

	We also record the near-field signal using headset microphones for each participant. To obtain the transcription, we first align the signal recorded by headset microphone and the first channel of microphone array and then select the signal with higher quality for manual labeling. Before labeling the scripts, we apply automatic speech recognition on the recorded data to assist transcribers to
	for accurate labeling results. Then inspectors will double-check the labeling results of each session from transcribers and decide whether meet the acceptance standard. Praat\footnote{https://github.com/praat/praat} is used for further calibration to check the accuracy of speaker distribution and to avoid the miss cutting of speech segments. We also pay special attention to accurate punctuation labeling. Each session is labeled by three professional annotators on average as a secondary inspection to improve the labeling quality.
	
	All scripts of the speech data are prepared in TextGrid format for each session, which contains the information of the session duration, speaker information (number of speaker, speaker-id, gender, etc.), the total number of segments of each speaker, the timestamp and transcription of each segment, etc. The non-speech events are transcribed as well, such as pauses, laughing, coughing, breathing, etc. The overlapping and non-overlapping segments are also identified.

	\begin{table}[htb]
		\caption{Session-level overlap ratio distribution. 0\%-10\%, 10\%-20\%, 20\%-30\%, 30\%-40\% and 40\%-100\% indicate the range of overlap ratio of each session. The numbers followed indicate the number of sessions with corresponding overlap ratio in training and evaluation sets, respectively.}
		
		\label{table:distribution}
		\centering
		\begin{tabular}{c|c|c}
			\toprule
			\hline
			Overlap Ratio  & Training & Evaluation \\ 
			\hline
			0\%-10\% & 41 & 12 \\
			10\%-20\% & 76 & 6 \\
			20\%-30\% & 44 & 2 \\
			30\%-40\% & 20 & 0 \\
			40\%-100\% & 10 & 0 \\
			\hline
			\bottomrule
		\end{tabular}
	\end{table}

	\section{Baseline}
	
	We build our baseline meeting transcription system\footnote{https://github.com/felixfuyihui/AISHELL-4}. inspired by~\cite{chen2020continuous, wang2021exploring}. The baseline system contains three major sub-modules, named speech front-end processing, speaker diarization and ASR. We train each sub-module separately. 
	During evaluation, the meeting recording is firstly processed by the speaker diarization module, which generates the speaker attribution and boundary information of each individual utterance. Then speech separation module is applied to each segmented utterance to remove the potential speech overlap. And finally, the processed audio is recognized by the ASR module to obtain the speaker attributed transcription.


	\subsection{Speaker Diarization}
	
	We adopt the Kaldi-based diarization system from CHiME-6 challenge in our system. The diarizaiton module includes three components: SAD, speaker embedding extractor and clustering. 
	
	We employ the CHiME-6's SAD~\cite{ghahremani2016acoustic} model\footnote{http://kaldi-asr.org/models/12/0012\_sad\_v1.tar.gz}, which consists of a 5-layer time-delay neural network (TDNN) and a 2-layer statistics pooling for continuous conference segmentation. The feature sent into the SAD model is 40 dimensional mel frequency cepstral coefficient (MFCC) extracted every 10 ms with 25 ms window. 
	
	The speaker embedding network \footnote{https://github.com/BUTSpeechFIT/VBx} is based on ResNet~\cite{zeinali2019but} which is trained using Voxceleb1~\cite{nagrani2017voxceleb}, Voxceleb2~\cite{nagrani2020voxceleb} and CN-Celeb~\cite{fan2020cn}. The speaker embedding network is trained using stochastic gradient descent (SGD) and additive angular margin loss~\cite{deng2019arcface}. We ramp-up the margin during the first two epochs and then train the network for the following epoch with fixed margin $m$ = 0.2. Data augmentation is performed in the same way as the SRE16 Kaldi recipe~\footnote{https://github.com/kaldi-asr/kaldi/blob/master/egs/sre16/v2}. 
	The feature fed into the speaker embedding network is 64 dimensional mel-filterbank which is extracted every 15 ms with 25 ms window. The post processing result of speaker embedding is derived from probabilistic linear discriminant analysis (PLDA)~\cite{ioffe2006probabilistic} model which is trained on the same data as the speaker embedding network. 
	
	All speaker embeddings generated will be pre-clustered using agglomerative hierarchical cluster (AHC)~\cite{han2008strategies} algorithm to obtain the initial speaker labels.
	The threshold of AHC algorithm is set to 0.015. The dimension of speaker embedding is reduced from 256 to 128 for further clustering using the Variational Bayesian HMM clustering (VBx)~\cite{landini2020bayesian} model. 
	Thus, the segmented utterances from the whole conference session together with their corresponding speaker information.
	
	Note that the speaker diarization does not take overlapped speech into consideration as we assumed that each speech frame corresponds to only one of the speakers. Diarization systems from~\cite{xiao2020microsoft} are designed to address speech overlap. We will explore their integration in follow-up works.
	

	\subsection{Speech Front-end Processing}
	
	We employ the separation solution from~\cite{yoshioka2019advances} in our baseline system, which contains a multi-channel speech separation network, and a mask based minimum variance distortionless response (MVDR) beamformer. For each testing segment, two masks are firstly estimated by the network, followed by the MVDR beamforming to generate the final separation output.
	
	The separation network consists of 3 LSTM layers, each having 3084 nodes, followed by a fully connection (FC) layer with sigmoid activation function to form speaker masks.
	
	We use synthetic overlapped speech to train the separation network. Each mixture sample consists of fully or partially overlapped speech, directional noise and isotropic noise.
	460 hours clean speech from LibriSpeech is adopted as the close-talk data for simulation. The noise set comes from MUSAN and Audioset which contain 49 and 88 hours of utterances, respectively. The multi-channel RIRs and isotropic noise are simulated based on the topology of 8-channel microphone array. The room size ranges from $3\times3\times3$ to $10\times10\times3$ m$^{3}$ while the microphone array is randomly placed within a $2\times2$ m$^{2}$ area in the center of the room with height randomly placed between 0.6 to 1.2 m. RT60 ranges from 0.2 to 0.8 s. The angle between each speaker and noise is set at least $20^{\circ}$ to ensure the spatial distinctiveness of each sound source. The SNR range is [5, 20] dB and SDR range is [-5, 5] dB. The overlap ratio of simulated data is divided into three parts with equal data amount: 1) non-overlapped, 2) overlap ratio randomly ranges from 0 to 20~\% and 3) overlap ratio randomly ranges from 20 to 80~\%. Finally we simulate 364 and 10 hours data for model training and development, respectively. 
	
	All simulated data is segmented into 4.0 s short utterances during training. Interchannel phase difference (IPD) features are calculated among four microphone pairs: (1,5), (2,6), (3,7), (4,8). The frame length and hop length used in STFT are set to 32 and 16 ms, respectively. The mask estimator is trained for 20 epochs with Adam optimizer, where the initial learning rate is set to 0.001 and will halve if there is no improvement on development set.
	
	We concatenate magnitude of the first channel of the observed signal, as well as the cosIPD among selected microphone pairs along frequency axis, as the input of the model. The IPD is calculated as $\text{IPD}_{i,j}=\angle\mathbf{y}_{t,f}^i-\angle\mathbf{y}_{t,f}^j$, where $i,j$ denote microphone index. For mask estimator, we use STFT mask as the training target and the loss function for magnitude approximation with the permutation invariant training (PIT) criteria is used for training.
	
	In MVDR step, 
	based on estimated masks, the covariance matrices are calculated via
	
	\begin{equation}
		\mathbf{R}_f^k = \frac{1}{\sum_t m_{t,f}^k} \sum_t m_{t,f}^k \mathbf{y}_{t,f} \mathbf{y}_{t,f}^\mathsf{H}.
	\end{equation}
	$m_{t,f}^k$ refers to the mask value of the source $k$ on each TF bin and $\mathbf{y}_{t,f}$ denotes the corresponding observed vector on the frequency domain. In this work, $k \in \{0, 1\}$ as we consider at most two speakers. The filter coefficients of the MVDR $\mathbf{w}_f^k$ for source $k$ is derived as follows
	\begin{equation}
		\mathbf{w}_f^k = \frac{(\mathbf{R}_f^{1-k})^{-1} \mathbf{R}_f^{k}}{\text{tr} \left((\mathbf{R}_f^{1-k})^{-1} \mathbf{R}_f^{k} \right)} \mathbf{u}_r,
	\end{equation}
	where $\text{tr}(\cdot)$ is the matrix trace operation and $\mathbf{u}_r$ is a one-hot vector indicating the reference microphone. Thus, we can get the final enhanced and separated result for each source via
	\begin{equation}
		\hat{\mathbf{y}}^{k}_{t,f}= {\mathbf{w}_f^{k}} ^{\mathsf{H}} \mathbf{y}_{t,f}.
	\end{equation}
	
	From the two separated channel of each utterance, we pick the channel with larger energy as the final processing result for this utterance.

	\subsection{Automatic Speech Recognition}
	
	A transformer based end to end speech recognition model ~\cite{dong2018speech} with sequence to sequence architecture is employed as the backend in our system.
	Here, a 2-layer 2D CNN is firstly applied to reduce the input frame rate, followed by a 8-layer transformer encoder and 6-layer transformer decoder. The multi-head self-attention layer contains 8 heads with 512 dimensions and the inner dimension of the feed forward network is 1024. The model is trained with joint CTC and cross-entropy objective function~\cite{watanabe2017hybrid} to accelerate the convergence of the training. The beam size is set to 24 during the beam search decoding.
	
	The training data is composed of two parts: simulated and real-recorded data. For simulated data, the clean speech data comes from AISHELL-1, aidatatang\_200zh and Primewords~\footnote{http://www.openslr.org/47/}, while noise, RIRs and SNR are the same as that of the front-end model. We combine the simulated and original data to form 768 and 97 hours single-speaker single-channel data as training and development set, respectively. For real-recorded data, we select the non-overlapped part of the training set of AISHELL-4 according to the ground-truth segmentation information, which consists of 63 hours data for model training only. 
	
	The frame length and hop size of STFT are set to 25 and 10 ms. We utilize 80 dimensional log-fbank features with utterance-wise mean variance normalization as the input feature and we apply SpecAugment ~\cite{park2019specaugment} for data augmentation. We train the model for a maximum of 50 epochs with the Adam optimizer. The warm-up and decay learning rate scheduler with a peak value of $10^{-4}$ is adopted, where the warm-up steps is set to 25,000.
	
	
	\subsection{Evaluation and Results}
	
	Two metrics are included in the baseline system, referring as speaker independent and speaker dependent character error rate (CER). The speaker independent task is designed to measure the performance of front-end and speech recognition module. In this task, we use the ground truth utterance boundary instead of the diarizaiton output to segment the utterance. On the other hand, the speaker dependent task takes consideration of all involved modules.
	After obtaining transcription or speaker attributed transcription, Asclite2\footnote{https://github.com/usnistgov/SCTK}, which can align multiple hypotheses against multiple reference transcriptions, is used to estimate CERs.

	We provide CER for reference system, with and without front-end, in Table~\ref{table:results}. Here, the CER results of no front-end processed data on speaker independent and speaker dependent tasks are 32.56\% and 41.55\%, respectively. While after frond-end processing, we get 30.49\% CER on speaker independent task, and 39.86\% CER on speaker dependent task when applying SAD and speaker diarization processing. In the future, we will consider deliver speech front-end processing in a continuous manner and reduce the usage of ground-truth information to get more promising results.
	\begin{table}[]
		\caption{\%CERs results for speaker independent and speaker dependent tasks.}
		\centering
		\label{table:results}
		\begin{tabular}{c|c|c}
			\toprule
			\hline
			& Speaker Independent & Speaker Dependent \\ \hline
			No FE & 32.56 &41.55 \\ \hline
			FE & 30.49 & 39.86\\
			\hline
			\bottomrule
		\end{tabular}
	\end{table}
	
	\section{Conclusions}
	In this paper, we present AISHELL-4, a sizable Mandarin speech dataset for speech enhancement, separation, recognition and speaker diarization in conference scenario. All data released are real recorded, multi-channel data in real conference venues and acoustic scenario. We also release the training and evaluation framework as baseline for research in this field. Our experiment results show that the complex acoustic scenario degrades the performance of ASR to a great extend, which results in 32.56\% and 41.55\% CERs in speaker independent and speaker dependent task, respectively. After the front-end and diarization processing, the CERs achieve 30.49\% and 39.86\% on speaker independent and speaker dependent task, respectively. We hope that this dataset and the associated baseline can promote researchers to explore different aspects in meeting processing tasks, ranging from individual tasks such as speech front-end processing, speech recognition and speaker diarization, to multi-modality modeling and joint optimization of relevant tasks, and help to reduce the gap between the research state and real conferencing applications.

	
	\bibliographystyle{IEEEtran}
	\bibliography{mybib}

	
\end{document}